\documentclass[%
 reprint,
superscriptaddress,
 amsmath,amssymb,
 aps,
]{revtex4-2}

\usepackage{graphicx}
\usepackage{dcolumn}
\usepackage{bm}
\usepackage{hyperref}

\newcommand{\TC}{$T_{\rm{c}}$}
\newcommand{\STO}{SrTiO$_3$}
\begin{document}

\preprint{APS/123-QED}

\title{Capping layer influence and isotropic in-plane upper critical field of the superconductivity at the FeSe/\STO~interface}
\author{Yanan Li}
\affiliation{Department of Physics, The Pennsylvania State University, University Park, PA 16802, USA}
\affiliation{International Center for Quantum Materials, School of Physics, Peking University, Beijing 100871, China}
\author{Ziqiao Wang}
\author{Run Xiao}
\author{Qi Li}
\affiliation{Department of Physics, The Pennsylvania State University, University Park, PA 16802, USA}
\author{Ke Wang}
\affiliation{Materials Research Institute, The Pennsylvania State University, University Park, PA 16802, USA}
\author{Anthony Richardella}
\affiliation{Department of Physics, The Pennsylvania State University, University Park, PA 16802, USA}
\affiliation{Materials Research Institute, The Pennsylvania State University, University Park, PA 16802, USA}
\author{Jian Wang}
\affiliation{International Center for Quantum Materials, School of Physics, Peking University, Beijing 100871, China}
\author{Nitin Samarth}
\thanks{Corresponding author: nsamarth@psu.edu}
\affiliation{Department of Physics, The Pennsylvania State University, University Park, PA 16802, USA}

\date{\today}

\begin{abstract}
Understanding the superconductivity at the interface of FeSe/\STO~is a problem of great contemporary interest due to the significant increase in critical temperature (\TC) compared to that of bulk FeSe, as well as the possibility of an unconventional pairing mechanism and topological superconductivity. We report a study of the influence of a capping layer on superconductivity in thin films of FeSe grown on \STO~using molecular beam epitaxy. We used {\it in vacuo} four-probe electrical resistance measurements and {\it ex situ} magneto-transport measurements to examine the effect of three capping layers that provide distinctly different charge transfer into FeSe: compound FeTe, non-metallic Te, and metallic Zr. Our results show that FeTe provides an optimal cap that barely influences the inherent \TC~found in pristine FeSe/\STO, while the transfer of holes from a non-metallic Te cap completely suppresses superconductivity and leads to insulating behavior. Finally, we used {\it ex situ} magnetoresistance measurements in FeTe-capped FeSe films to extract the angular dependence of the in-plane upper critical magnetic field. Our observations reveal an almost isotropic in-plane upper critical field, providing insight into the symmetry and pairing mechanism of high temperature superconductivity in FeSe.
\end{abstract}

\maketitle
\section{\label{sec:level1}Introduction}

The study of superconductivity in single unit cell FeSe interfaced with TiO$_2$ terminated \STO~(STO) is largely motivated by the discovery of intriguingly high critical temperatures. While \TC  $\sim 9.4$ K for bulk FeSe \cite{Hsu14262} and \TC $\sim 3.7$ K in two-layer FeSe films grown on bilayer graphene \cite{PhysRevB.84.020503}, {\it in situ} scanning tunneling spectroscopy (STS) measurements of single unit cell FeSe/STO and angle resolved photo emission spectroscopy (ARPES) measurements show a gap closing critical temperature around 65 K \cite{Wang_2012,Liu:2012aa,He2013,Peng2014}; one report of {\it in situ} transport measurements (still to be reproduced) even suggests the \TC~might reach 100 K in this system \cite{Ge:2015aa}. This material system thus provides an exciting platform to investigate high \TC~superconductivity different from that in the more extensively studied cuprates. It has also been suggested to provide a pathway to achieve topological superconductivity \cite{Wang:2016aa,Wang_2017,Liueaax7547,Chen:2020aa,Liu_2020}. 

In order to carry out {\it ex situ} transport measurements on these ultrathin FeSe films, a typical approach has used protective capping layers such as FeTe and amorphous Si. An onset \TC~is detected around 54.5 K for {\it ex situ} measurements \cite{ZHANG:17401}, slightly lower than the value given by {\it in situ} STS measurements. In order to seek a better understanding of the role played by the protection layer, we conducted {\it in situ} and {\it ex situ} transport measurements on FeSe films grown on STO with various capping layers. By using {\it ex situ} Hall measurements, we studied the relationship between the charge transfer from the capping layer and its influence on the superconductivity in FeSe thin films. Our results indicate that a compound capping layer (FeTe) and a metallic capping layer (Zr) have little influence on \TC. (The FeTe reduces \TC~slightly, by $\sim 2$ K). In strong contrast, when we use a Te capping layer, superconductivity is completely suppressed and a superconductor to insulator transition occurs at low temperature. This is likely due to the transfer of holes from the Te layer to interfacial FeSe. 
We also used the FeSe/STO film with crystalline FeTe capping to study the angle dependent in-plane upper critical field. This allowed us to study the pairing mechanism in FeSe. Previous experimental and theoretically investigations attribute the high \TC~gap to the cooperative effect of spin fluctuations, band bending-induced charge transfer from STO, and interfacial electron-phonon interaction \cite{DHLee_2018,Zhaoeaao2682,Song:2019aa,Zhang:2017aa,Lee:2014aa}. In addition, $s\pm$ wave superconductivity has been considered to be a likely candidate for the gap pairing symmetry in iron pnictide superconductors, where the pairing ``glue” is mediated by spin fluctuations \cite{PhysRevLett.101.057003}. However, it is still debated whether $s\pm$ wave can describe the superconductivity in FeSe/STO, where there is no hole band on the Fermi surface and the STO phonon mode likely participates in the pairing. In fact, various pairing symmetries have been proposed in this system, including plain $s$ wave, $s{++}$, $s\pm$ wave, and $d$ wave \cite{Fan:2015aa,Ge:2019aa,Chen_PhysRevLett.124.097001,CLiu_PhysRevLett.123.036801,Agterberg_PhysRevLett.119.267001,Yamakawa_PhysRevB.96.045130,Hu_PhysRevX.3.031004,Chen_PhysRevB.92.224514,Yang_PhysRevB.88.100504,Kang_PhysRevLett.117.217003,doi:10.1021/acs.nanolett.9b00144}. As the angle dependence of the upper critical magnetic field ($H_{c2}$) can shed light on the gap symmetry, we carried out angle-dependent in-plane ({\it ab} plane) magneto-resistance measurements. Our observations support an almost isotropic in-plane $H_{c2}$ below \TC. We discuss this in relation to the proposed paring symmetries and other effects that might mask an anisotropic gap, if one exists. Our findings provide a basis for further studies of the pairing symmetry of the interface superconductivity of FeSe/STO in the future. 

\section{\label{sec:level2}Experimental Methods}
We grew FeSe films on TiO$_2$-terminated STO (001) substrates with atomically flat steps in a Scienta Omicron molecular beam epitaxy (MBE) system. The base pressure of our system is $10^{-11}$ torr. The Fe, Se, and Te were evaporated from thermal Knudsen effusion cells. Zr was evaporated using an electron beam evaporator at 7 kV and 100 mA. To obtain TiO$_2$-terminated STO substrates, we first  pretreated commercial (Shinkosha, Japan) substrates by standard chemical etching with 90$^{\circ}$C deionized water (45 mins) and 10$\%$ HCl solution (room temperature, 45 mins). Then, we annealed the pretreated substrates in a tube furnace under oxygen flow at 980$^{\circ}$C for 3h \cite{ZHANG:17401,Zhaoeaao2682}. FeSe films with thickness between 3-7 unit cells were then grown using the procedures described by Zhang {\it et al} \cite{ZHANG:17401}. During the growth, the substrate was kept at 450$^{\circ}$C. The growth rate is around 0.15 layer/min. After growth, the sample was annealed at 500$^{\circ}$C for 2h. Zr and Te capping layers were deposited at a substrate temperature $T_S = 20^{\circ} \rm{C}$ (as measured using a thermal imaging camera). The FeTe caps were deposited at $T_S = 350^{\circ} \rm{C}$. The MBE system is connected {\it in vacuo} with a Scienta Omicron LT-Nano four-probe scanning tunneling microscope (STM) that allows four scanning tips to move independently on the sample surface. The {\it in situ} transport measurements were conducted in this system without exposing the films to air, and the distance between the tips is around 2 mm during the measurement. The {\it ex situ} transport measurements were conducted in a Quantum Design 9 T Physical Property Measurement System (PPMS), and the film samples are mechanically scratched into Hall bar devices for {\it ex situ} transport measurements.

Cross-sectional transmission electron microscope (TEM) specimens were prepared on an FEI Helios 660 focused ion beam (FIB) system. 1 kV final cleaning was applied after samples became electron transparent to avoid ion beam damage to the sample surfaces. The high angle annular dark field (HAADF) scanning transmission electron microscopy (STEM) was performed on an FEI STEM at 200 kV (Titan G2 60-300). Energy-dispersive X-ray spectroscopic (EDS) elemental maps were collected by using a SuperX EDS system (Bruker) under STEM mode.

\section{\label{sec:level3}Results and discussion}

\begin{figure}
\includegraphics[width=0.4\textwidth]{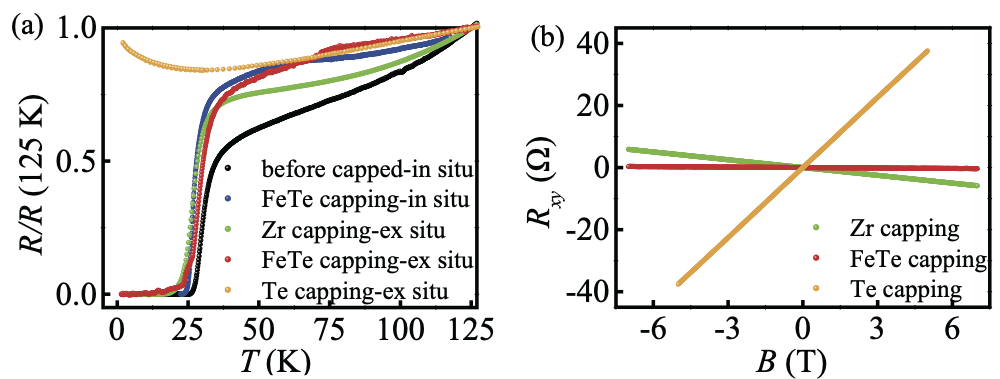}
\caption{\label{FIG.1} Capping layer influence of superconductivity in FeSe films grown on STO. (a) Normalized temperature-dependent resistance of a FeSe film from {\it in situ} and {\it ex situ} electrical transport measurements. The FeTe capping layer decreases the \TC~of FeSe by around 2 K. All samples have FeSe thickness in the range 3-4 unit cells. (b) {\it Ex situ} measurements of  Hall resistance of an FeSe film with a Zr cap ($n=6.9\times10^{18}\ \mathrm{m}^{-2}$), a FeTe cap  ($n=1.25\times10^{20}\ \mathrm{m}^{-2}$), and a Te cap  ($p=8.3\times10^{17}\ \mathrm{m}^{-2}$), respectively. The FeTe capped sample has FeSe thickness of 7.5 UC.  The other two samples have FeSe thickness in the range 3-4 unit cells. The data are anti-symmetrized in magnetic field.
}
\end{figure}
\subsection{\label{sec:levela}Capping layer influence}

Figure 1(a) shows the temperature dependence of the four-probe resistance of an FeSe film before and after capping with various layers (FeTe, Zr, Te). The resistance is normalized with respect to the value at 125 K. {\it In situ} measurements show that the zero-resistance \TC~of the bare FeSe film is around 26 K. After capping with FeTe, this {\it in situ} \TC~decreases slightly to 24 K. Earlier work has shown that post-growth annealing at a relatively high temperature is an essential step for achieving high \TC~superconductivity in single unit cell FeSe/STO. It is speculated that this annealing removes excess Se and allows the transfer of electrons from STO into FeSe \cite{Wang_2017-1,He18501}. When we cap the pristine FeSe layer with FeTe, the Te flux is usually several times and even ten or twenty times larger than the Fe flux. The growth of FeTe introduces surplus Te onto the surface of FeSe/STO, introducing extra holes into the film, which might slightly suppress the \TC. Additionally, the growth of capping layers is likely to introduce impurities and defects into the film, which could also result in the decrease of the  \TC. It should be noted that the decrease in \TC~caused by the overgrowth of FeTe is small (only $\sim 2$ K). 

From 125 K down to low temperature before superconducting transition, the resistance of the film with a capping layer decreases slower than that of the film without a capping layer. All these three types of capping layers decrease the residual resistivity ratio (RRR) of the FeSe film, indicating that all these capping layers likely introduce impurities and defects into the FeSe film. Figure 1(b) show the Hall resistance of FeSe films capped with FeTe, Zr, and Te from {\it ex situ} transport measurements. It is measured at 50 K for FeTe capping and Zr capping, and at 30 K for Te capping. Both the samples capped with FeTe and Zr show a superconducting transition at low temperature, and transport is dominated by n-type carriers (electrons). The FeSe film capped with only Te is not superconducting, but displays an insulating transition at low temperature even though it shows metallic behavior at higher temperatures. Charge transfer from the Te capping layer changes the dominant carriers in the FeSe film from electrons into holes. This shows a capping layer induced insulator transition in FeSe film, similar to the insulator transition in as grown p-type FeSe films or p-type FeSe films that have been annealed for a short time post-growth \cite{Wang_2017-1,He18501}. 

Atomic resolution HAADF STEM measurements indicate that the Zr capping layer (Fig. 2(a)) is amorphous and that the crystallinity of the FeSe film is also slightly destroyed by the capping process, although a superconducting transition is still observed by transport measurements.The circled area in the TEM image does not have as high crystalline quality as other areas. We tested four samples with Zr capping. Some samples have lower \TC, and the crystalline structure at the interface in these samples is disrupted even more severely by the capping process, as shown by TEM results (Fig. S1 in Supplementary Material \cite{Supplementary}). It is likely that the high energy Zr particles evaporated by the electron beam create damage when they impinge on the FeSe film. Growing the Zr capping layer at a much lower substrate temperature or lower e-beam power might be helpful to avoid this issue. Similar to FeTe capping, Te forms a crystalline capping layer as shown in Fig. 2(c). The electron dispersive spectroscopy (EDS) oxygen mapping results (Fig. 2(b)) show that Zr capping can capture oxygen and stop further oxidation of the FeSe film while Te protects the film by resisting the penetration of Oxygen (Fig. 2(d)). Possibly the Zr oxidizes initially, and after that oxygen doesn’t diffuse through it easily. EDS mapping of the elemental distribution in an FeSe film capped with Zr and Te (Figs. S2 and S3 Supplementary Material \cite{Supplementary}) shows that the interface between Zr/Te and FeSe is clean: we do not detect Zr or Te signals in the FeSe film within the EDS resolution sensitivity limits. The interface between the FeSe film and crystalline FeTe and Te capping layers is cleaner than that between FeSe and a Zr capping layer. 

Taking the transport results into consideration as well, this allows us to conclude that non-metallic Te does not provide a good capping layer for protecting the thin FeSe film for {\it ex situ} experimental studies of its superconducting properties. Nonetheless, we note that an FeSe film capped with this kind of non-metallic material can recover the superconducting transition after being de-capped and annealed again at high temperature {\it in vacuo} \cite{Liu:2012aa}.  Earlier studies have shown that that electron doping on the surface of a monolayer FeSe film on STO by K cannot further increase the \TC~beyond the enhancement already obtained by interfacing with STO \cite{Zhang:2016aa}, which again suggests that the charge transfer and the phonon mode from STO work cooperatively to enhance the \TC~of this system. To further enhance the \TC~with a capping layer, some oxide capping layer which can transfer electron charge and contribute phonon mode into the film is desirable. Our results demonstrate that the effect of the charge transfer from a capping layer can play an important role in the superconductivity of FeSe/STO. Understanding the influence of the capping layer is necessary in order to perform the {\it ex situ} measurements, especially if a consistent result between {\it in situ} and {\it ex situ} measurements is expected. Having gained some insight into the influence of the capping layer on superconductivity, we now discuss our attempts to understand the pairing symmetry in FeSe/STO. We use FeTe capping as the protective layer on top of FeSe for this purpose.

\begin{figure}
\includegraphics[width=0.5\textwidth]{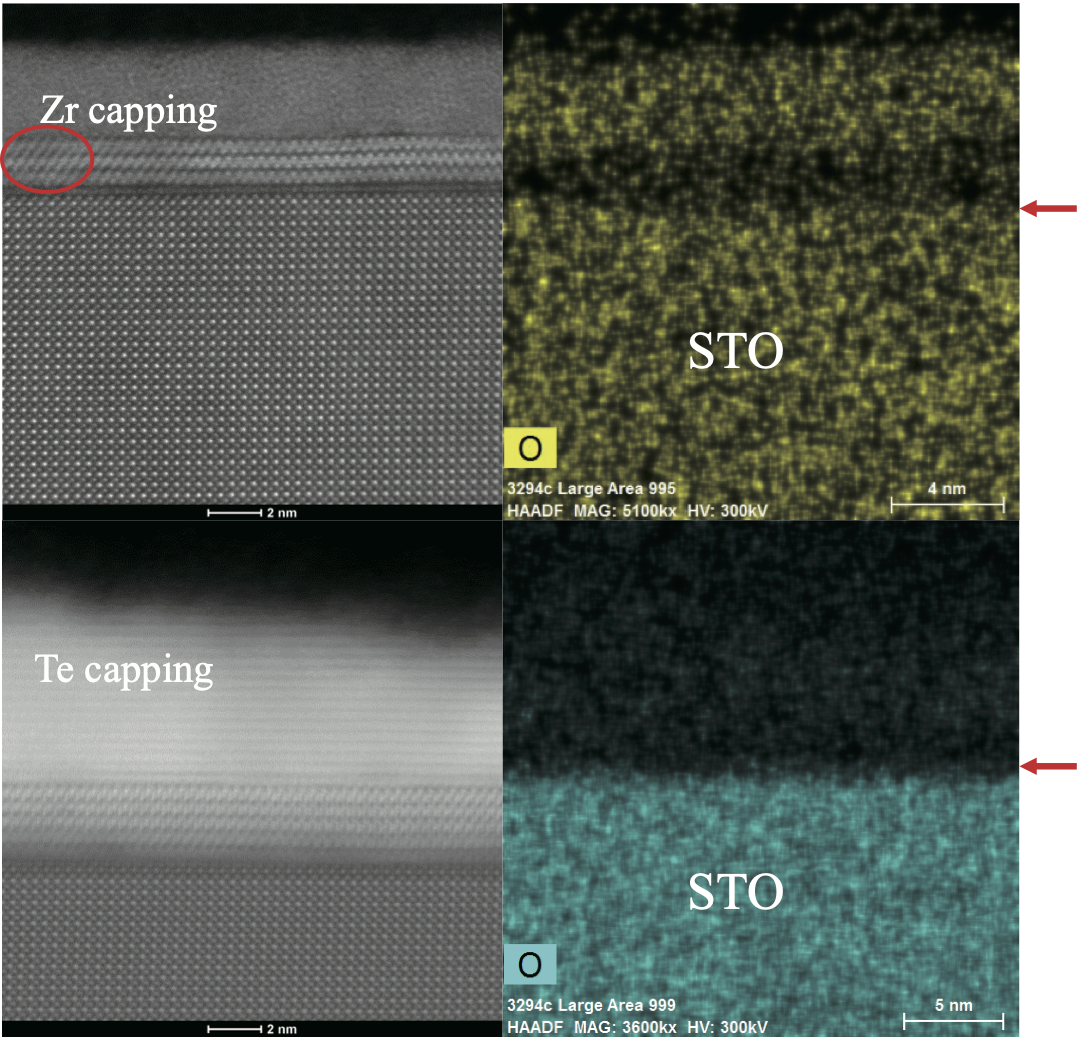}
\caption{\label{FIG.2} Atomic resolution HAADF STEM images (a and c) and EDS mapping (b and d) of oxygen of FeSe film with Zr and Te. (a and b) FeSe capped with Zr, (c and d) FeSe capped with Te. The EDS results are on the same scale and the same region as the corresponding STEM images. The Zr layer has an amorphous structure and it partially destroys the crystalline structure of FeSe. The red-circled area does not have as high crystalline quality as other areas. The blue arrows indicate the interface between FeSe film and STO. The red arrows indicate the interface between FeSe film and the capping layer.   
}
\end{figure}

\subsection{\label{sec:levelb}Isotropic in-plane upper critical field}

We carried out angle dependent in-plane magneto-resistance measurements in FeTe/FeSe/STO samples both below and above \TC~to search for anisotropy of the magneto-resistance in the superconducting region. The presence or absence of an observed anisotropy can provide important insights into the pairing symmetry. We caution that such a measurement has to be carried out with great care since a very small misalignment ($< 1^{\circ}$) between the plane of the sample and the magnetic field can easily result in a misleading artifact signal suggestive of an anisotropic magneto-resistance. Figure 3(a) shows the temperature dependence of the resistance, indicating the onset \TC~is around 45 K. Figure 3(b) shows the in-plane magneto-resistance at various temperatures. At first sight, we do indeed appear to observe an interesting anisotropy in the superconducting region which weakens with increasing temperature and disappears gradually above \TC. We obtained data similar to this in 10 distinct sample mountings and cool downs. However, we have concluded that this oft-observed data is an example of the artifact mentioned above. With a careful remounting of the sample accompanied by a rotation in the {\it ab} plane by $45^{\circ}$, the maximum and minimum angle of anisotropy of the magneto-resistance changed by a different value (not $45^{\circ}$) compared to the change of the sample orientation (see Fig. 3(c)). The amplitude also changed. As the sample was measured in the same instrument with the same setup for both orientations, these observations suggest that the anisotropy is more likely due to a small out-of-plane misalignment angle when mounting the sample rather than an intrinsic effect originating from the sample itself. To confirm this hypothesis, we plot the angle-dependence of the critical field at 25 K extracted from the magnetic field dependence of the resistance (Fig. S4 in Supplementary Material)\cite{Supplementary} at various angles in (Fig. 3(d)). This anisotropic $H_{c2}$ can be fit with the 2D Tinkham formula\cite{PhysRev.129.2413}:
\begin{equation}
\Big[\frac{H_{C2} (\theta') \sin (\theta')}{H_{C2}^{||}}\Big]^2 + \Big|\frac{H_{C2} (\theta') \cos (\theta')}{H_{C2}^{\perp}}\Big| = 1
\end{equation}
by considering a small misalignment angle $\alpha$ from the in-plane direction $\cos \theta ' = \cos (\theta - 70^{\circ}) \sin \alpha$ (see the insets of Fig. 3(c) and (d) for schematics of the angle positions $\theta$ and $\alpha$). Here, $\theta = 70^{\circ}$ (minimum of the upper critical field) is taken as the out-of-plane direction and $\theta = 160^{\circ}$ (maximum of the upper critical field) is taken as the in-plane direction. GE varnish is used for mounting the sample, and a small misalignment angle between sample and sample holder is usually unavoidable. As the misalignment angle is random in both direction and amplitude when we mount the sample, the observed anisotropic magneto-resistance has different peak position and oscillation amplitude. This ``artificial" anisotropy is absent above \TC~because the magneto-resistance amplitude is very small above \TC. If we exclude the anisotropy due to the misalignment angle, our observations support an almost isotropic in-plane upper critical field within the resolution limit of our instrument. 

\begin{figure*}
\includegraphics[width=0.8\textwidth]{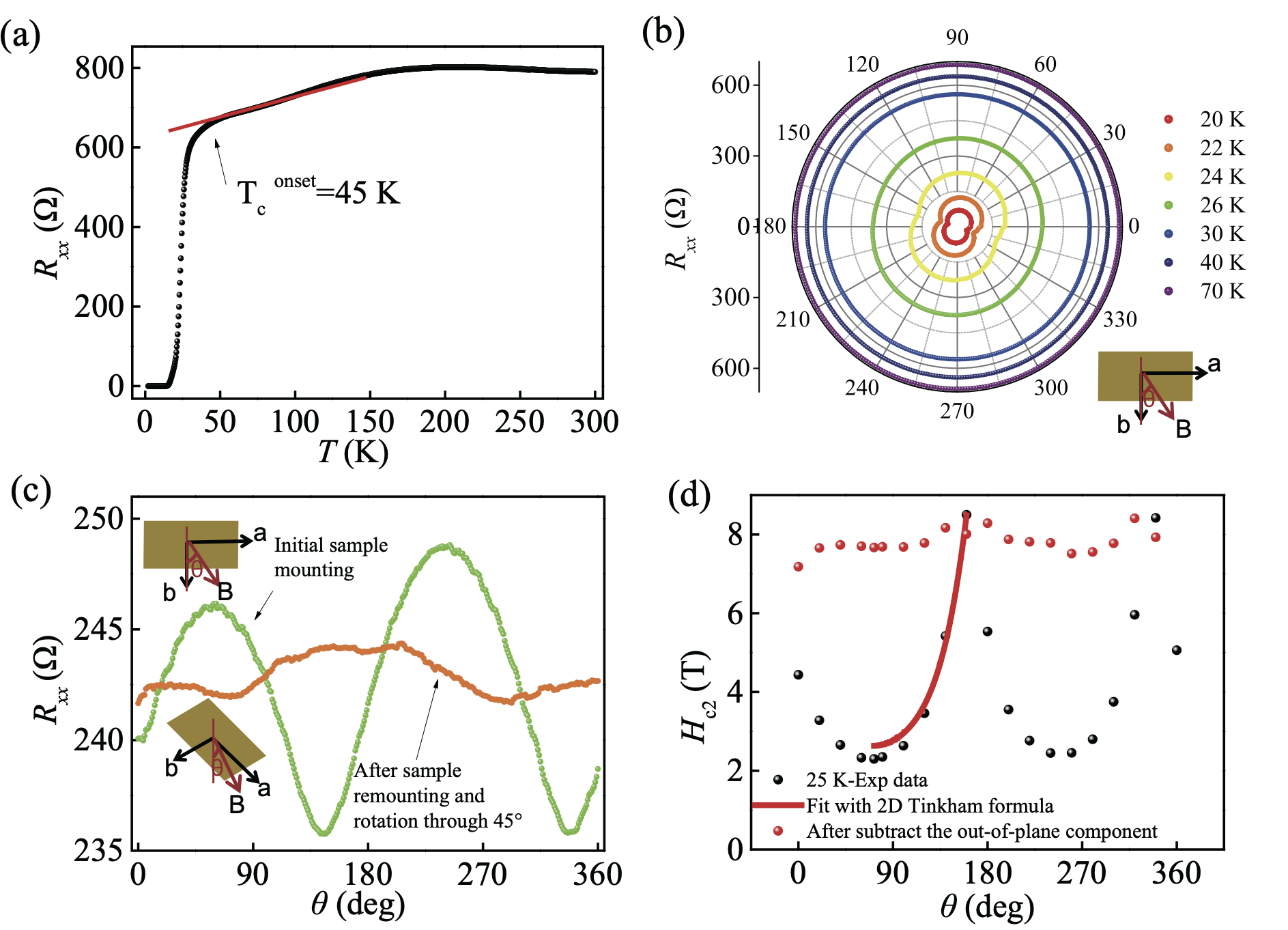}
\caption{\label{FIG.3} Absence of anisotropy of in-plane upper critical field of FeSe/STO. (a) Temperature dependent resistance curve of an FeTe-capped FeSe film, showing the superconductivity transition with onset \TC~around 45 K. The thickness of the film is estimated to be 3 unit cells. (b) Angle-dependence of the in-plane magnetoresistance at various temperatures. (c) Comparison of angle-dependence of the magnetoresistance after rotating and remounting the sample in plane through $45^{\circ}$ taken at 18 K and 9 T. The angle position of the two-fold symmetry shifts $70^{\circ}$, and the amplitude becomes smaller, which indicates the two-fold symmetry is not an intrinsic in-plane property of the FeSe film but rather arises from the out-of-plane misalignment. Insets are the schematics of the measurement setup. $\theta$ is the in-plane angle position, and $\alpha$ is the out-of-plane misalignment angle. The magnetoresistance is normalized with respect to the minimum value of the magnetoresistance. (d) Angle-dependence of the critical field extracted from resistance {\it vs} field curves at various angles at 25 K. The angle dependence (black dots are the experimental date) can be fit with the 2D Tinkham formula (red curve) by considering a fixed out-of-plane misalignment angle $\alpha$, confirming that the two-fold symmetry is from the out-of-plane misalignment. Therefore, the in-plane critical field is isotropic within the resolution limit of the measurement.
}
\end{figure*}

The upper critical field of a superconductor is determined by the interaction of an external applied magnetic field with the orbital and spin degrees of the charge in Cooper pairs via the orbital effect and Pauli paramagnetism  \cite{Wang_PhysRevB.94.014501}. When the orbital effect is considered, $H_{c2}$ is determined by the Fermi surface topology and the coupling parameter of the superconductor. Thus, the angular dependence of $H_{c2}$ can reveal insights into the pairing symmetry. For instance, some $d$-wave superconductors show four-fold oscillations in the in-basal-plane angle dependent $H_{c2}$ with the maximum along the antinodal direction and a minimum along the nodal direction \cite{Wang_PhysRevB.94.014501}. ARPES measurements show that the Fermi surface of interfacial FeSe/STO is composed of two ellipsoidal electron pockets overlapping with each other at the Brillouin zone corner (four M-points)\cite{Zhang_PhysRevLett.117.117001} (see Fig. S5 for a diagram of the Fermi surface).  The superconducting gap is nodeless but moderately anisotropic with the gap maxima located along the major axis of the ellipse and the minimum along the intersection of two ellipsoidal electron pockets \cite{Zhang_PhysRevLett.117.117001}. As mentioned earlier, theoretical proposals and experimental investigations have yet to arrive at a unified conclusion about the role of $s++$, hidden $s\pm$, extended $s\pm$, odd-parity $s-$, and nodeless $d$-wave pairing in FeSe. Amongst these possibilities, a nodeless $d$-wave pairing scenario is proposed when the relevant spin orbit coupling energy of FeSe/STO is smaller than the superconducting gap. One signature of this kind of $d$-wave pairing is that the gap minima and maxima will evolve in response to the external fields, temperature, or pressure\cite{Agterberg_PhysRevLett.119.267001}. Nevertheless, our measurements show that the in-plane upper critical field of FeSe/STO is isotropic within the resolution limit of the instrument. Thus, we do not detect any signature of a superconducting gap anisotropy. This observed isotropic in-plane $H_{c2}$ below \TC~could be explained by three scenarios. First, the isotropic in-plane magneto-resistance above \TC~ (Fig. 3 (b)) indicates that the scattering and spin orientation is isotropic. The anisotropy of the superconducting gap is weak, and possibly hidden by the isotropic scattering in the transport behavior. Second, the superconductivity in FeSe/STO lies in the Bardeen-Cooper-Schrieffer (BCS) to Bose-Einstein condensation (BEC) crossover regime, near the unitary point \cite{Zhang_CPL2019}. In the BCS state, the fermions pair into bosonic Cooper pairs with a weak attractive interaction while the particles in the BEC state are all tightly bound bosonic pairs. In the BCS limit, fermions pair up and condense at the same critical temperature. In contrast, the particles in the BEC state first pair up and then condense below \TC. Therefore, the evolution of critical parameters (\TC~and $H_{c2}$) does not reflect the gap symmetry when the superconductor is in the vicinity of the BEC state and the anisotropic superconducting gap does not result in an anisotropic critical field. Finally, the orbital effect might be strongly suppressed when the superconductivity is quasi two-dimensional (out-of-plane Fermi velocity is nearly zero). Consequently, the angle dependence of the in-plane $H_{c2}$ is mainly determined by the angular dependence of the spin susceptibility and spin orbital coupling of the sample and does not reflect the gap symmetry.

\section{\label{sec:level4}Conclusions}

To conclude, we studied the capping layer influence on superconductivity in FeSe/STO by combining {\it in situ} and {\it ex situ} electrical transport measurements. We found that an FeTe capping layer slightly decreases the \TC~of the FeSe film but otherwise preserves the superconducting properties for {\it ex situ} studies. Non-metallic capping layers such as Te are not suitable for {\it ex situ} experimental study of FeSe/STO as they transfer holes into FeSe, inducing an insulator transition. E-beam evaporated Zr does protect the FeSe film from oxidization and preserves the superconductivity, but it also can damage the FeSe  crystalline structure at the interface to some extent. Further explorations about metallic capping layer are desired. Finally, we used in-plane angle-dependent magneto-resistance measurements to demonstrate that the in-plane upper critical magnetic field of FeSe/STO is isotropic, and the possible origins are discussed.

\begin{acknowledgments}
This research was carried out using the Penn State Two-Dimensional Crystal Consortium-Materials Innovation Platform (2DCC-MIP) under NSF Grant No. DMR-1539916. We thank Cui-zu Chang and Timothy Pillsbury for their helpful suggestions in sample preparation. YL acknowledges support from the University of Chicago. AR and NS acknowledge support from NSF Grant No. DMR-1539916. RX acknowledges support from the Institute for Quantum Matter under DOE EFRC grant DE-SC0019331. QL acknowledges support from DOE under grant No. DE-FG02-08ER46531. JW acknowledges support from National Natural Science Foundation of China (No. 11888101). 
\end{acknowledgments}


\providecommand{\noopsort}[1]{}\providecommand{\singleletter}[1]{#1}%

\end{document}